\documentclass[10pt,twocolumn]{article}
\usepackage{fullpage}
\usepackage{epsfig}
\usepackage{times}

\long\def\com#1{}

\def\widowpage{\newpage}

\def\vxzip{vxZIP}
\def\vxunz{vxUnZIP}
\def\vx32{vx32}

\title{VXA: A Virtual Architecture for Durable Compressed Archives}
\author{Bryan Ford \\
	Computer Science and Artificial Intelligence Laboratory \\
	Massachusetts Institute of Technology}
\date{}

\begin{document}
\maketitle

\com{
\barenote{
This technical report describes a work in progress
and does not contain complete, final, or polished results.
This research was conducted as part of the IRIS project
({\tt http://project-iris.net/}),
supported by the National Science Foundation
under Cooperative Agreement No. ANI-0225660.}
}


\begin{abstract}
Data compression algorithms change frequently,
and obsolete decoders
do not always run on new hardware and operating systems,
threatening the long-term usability
of content archived using those algorithms.
Re-encoding content into new formats is cumbersome,
and highly undesirable when lossy compression is involved.
Processor architectures,
in contrast,
have remained comparatively stable
over recent decades.
VXA, an archival storage system designed around this observation,
archives executable decoders along with the encoded content it stores.
VXA decoders run in a specialized virtual machine
that implements an OS-independent execution environment
based on the standard x86 architecture.
The VXA virtual machine strictly limits access
to host system services,
making decoders safe to run
even if an archive contains malicious code.
VXA's adoption of a ``native'' processor architecture
instead of type-safe language technology
allows reuse of existing ``hand-optimized'' decoders
in C and assembly language,
and permits decoders access to
performance-enhancing architecture features
such as vector processing instructions.
The performance cost of VXA's virtualization is typically less than 15\%
compared with the same decoders running natively.
The storage cost of archived decoders,
typically 30--130KB each,
can be amortized across many archived files
sharing the same compression method.
\end{abstract}

\section{Introduction}
\label{sec-intro}

\begin{figure}[t]
\centerline{\epsfig{file=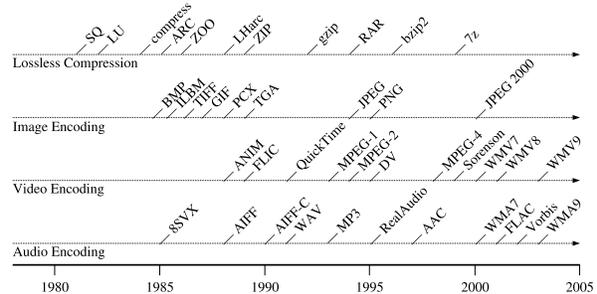, scale=0.22}}
\caption{Timeline of Data Compression Formats}
\label{fig-timeline}
\end{figure}

Data compression techniques
have evolved rapidly
throughout the history of personal computing.
Figure~\ref{fig-timeline} shows a timeline for the introduction of
some of the most historically popular compression formats,
both for general-purpose data and for specific media types.
(Many of these formats actually support
multiple distinct compression schemes.)
As the timeline illustrates,
common compression schemes change every few years,
and the explosion of lossy multimedia encoders
in the past decade
has further accelerated this evolution.
This constant churn in popular encoding formats,
along with the prevalence of other less common,
proprietary or specialized schemes,
creates substantial challenges to preserving the usability
of digital information over the long term~\cite{garrett96preserving}.

Open compression standards,
even when available and widely adopted,
do not fully solve these challenges.
Specification ambiguities
and implementation bugs
can make content encoded by one application
decode incorrectly or not at all in another.
Intellectual property issues such as patents
may interfere with the widespread availability of decoders
even for ``open'' standards,
as occurred in the last decade~\cite{battilana04gif}
with several file formats based on the LZW algorithm~\cite{nelson89lzw}.
Standards also evolve over time,
which can make it increasingly difficult to find
decoders for obsolete formats
that still run on the latest operating systems.

\begin{figure}[t]
\centerline{\epsfig{file=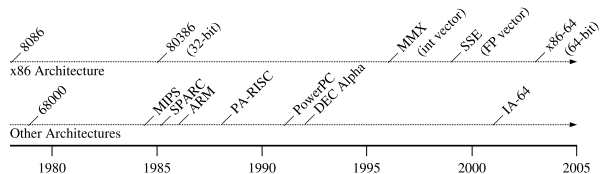, scale=0.22}}
\caption{Timeline of Processor Architectures}
\label{fig-archline}
\end{figure}

Processor architectures, in contrast,
have shown remarkable resistance to change
ever since the IBM PC first jump-started personal computing.
As the architecture timeline in Figure~\ref{fig-archline} illustrates,
the persistently dominant x86 architecture
has experienced only a few major architectural changes
during its lifetime---%
32-bit registers and addressing in 1985,
vector processing upgrades starting in 1996,
and 64-bit registers and addressing in 2003.
More importantly,
each of these upgrades has religiously preserved backward code compatibility.
Of the other architectures introduced during this period,
none have come close to displacing the x86 architecture in the mainstream.

From these facts we observe that
{\em instruction encodings are historically
more durable than data encodings}.
We will still be able to run x86 code efficiently
decades from now,
but it is less likely
that future operating systems and applications
will still include robust, actively-maintained decoders
for today's compressed data streams.

\subsection{Virtualizing Decoders}

{\em Virtual eXecutable Archives}, or VXA,
is a novel archival storage architecture
that preserves data usability
by packaging executable x86-based decoders
along with compressed content.
These decoders run in a specialized virtual machine (VM)
that minimizes dependence
on evolving host operating systems and processors.
VXA decoders run on
a well-defined subset
of the unprivileged 32-bit x86 instruction set,
and have no direct access to host OS services.
A decoder only extracts archived data
into simpler,
and thus hopefully more ``future-proof,'' uncompressed formats:
decoders cannot have user interfaces,
open arbitrary files,
or communicate with other processes.

\com{
The VXA virtual machine's deterministic execution model
guarantees that
a decoder always yields the same results
given the same input,
even in the presence of encoder or decoder bugs.
If a particular codec pair
does not quite conform to its intended encoding standard,
for example,
data compressed with the encoder and archived with the matching decoder
will still decode correctly forever,
if it decodes correctly when initially archived.
}

By building on the ubiquitous native x86 architecture
instead of using a specialized abstract machine
such as Lorie's archival
``Universal Virtual Computer''~\cite{lorie00archiving},
VXA enables easy re-use of existing decoders
written in arbitrary languages such as C and assembly language,
which can be built with familiar development tools such as GCC.
Use of the x86 architecture
also makes execution of virtualized decoders
extremely efficient on x86-based host machines,
which is important to the many popular ``short-term'' uses of archives
such as backups, software distribution, and structured document compression.
VXA permits decoders
access to the x86 vector processing instructions,
further enhancing the performance
of multimedia codecs.

Besides preserving long-term data usability,
the VXA virtual machine also isolates the host system
from buggy or malicious decoders.
Decoder security vulnerabilities,
such as the recent critical JPEG bug~\cite{msjpegbug},
cannot compromise the host under VXA.
This security benefit is important
because data decoders tend to be inherently complex
and difficult to validate,
they are frequently exposed to data
arriving from untrusted sources such as the Web,
and they are usually perceived as too low-level and performance-critical
to be written in type-safe languages.

\subsection{Prototype Implementation}

A prototype implementation of the VXA architecture,
\vxzip{}/\vxunz{},
extends the well-known ZIP/UnZIP archive tools
with support for virtualized decoders.
The \vxzip{} archiver can attach VXA decoders
both to files it compresses
and to input files already compressed
with recognized lossy or lossless algorithms.
The \vxunz{} archive reader
runs these VXA decoders to extract compressed files.
Besides enhancing the durability of ZIP files themselves,
\vxzip{} thus also enhances the durability
of pre-compressed data stored in ZIP files,
and can evolve to employ the latest specialized compression schemes
without restricting the usability of the resulting archives.

VXA decoders stored in \vxzip{} archives are themselves compressed
using a fixed algorithm
(the ``deflate'' method standard for existing ZIP files)
to reduce their storage overhead.
The \vxzip{} prototype currently includes six decoders
for both general-purpose data and specialized multimedia streams,
ranging from 26 to 130KB in compressed size.
Though this storage overhead may be significant for small archives,
it is usually negligible for larger archives
in which many files share the same decoder.

The prototype \vxzip{}/\vxunz{} tools
run on both the 32-bit and 64-bit variants of the x86 architecture,
and rely only on unprivileged facilities
available on any mature x86 operating system.
The performance cost of virtualization,
compared with native x86-32 execution,
is between 0 and 11\%
measured across six widely-available general-purpose and multimedia codecs.
The cost is somewhat higher, 8--31\%,
compared with native x86-64 execution,
but this difference is due not to virtualization overhead
but to the fact that VXA decoders are always 32-bit,
and thus cannot take advantage of the new 64-bit instruction set.
The virtual machine that \vxunz{} uses to run the archived decoders
is also available as a standalone library,
which can be re-used
to implement virtualization and isolation of extension modules
for other applications.

\com{
Implementing the VXA VM's fully deterministic execution model
presents a challenge even on x86-based host processors,
because x86 instructions often yield architecturally undefined,
and thus processor implementation-dependent,
values in condition codes or corner-case results.
The VM must therefore ``nail down'' these behaviors
by dynamically re-writing the instruction streams of VXA decoders.
Several simple optimizations
nevertheless keep the cost of this instruction translation
down to less than a XXX\%
compared to native execution.
}

Section~\ref{sec-arch} of this paper
first presents the VXA architecture in detail.
Section~\ref{sec-impl} then describes
the prototype \vxzip{}/\vxunz{} tools,
and Section~\ref{sec-vm} details
the virtual machine monitor
in which \vxunz{} runs archived decoders.
Section~\ref{sec-eval} evaluates
the performance and storage costs of the virtualized decoders.
Finally,
Section~\ref{sec-related} summarizes related work,
and Section~\ref{sec-conc} concludes.

\section{System Architecture}
\label{sec-arch}

This section introduces
the {\em Virtual eXecutable Archive} (VXA) architecture
at a high level.
The principles described in this section
are generic and should be applicable
to data compression, backup, and archival storage systems of all kinds.
All implementation details specific
to the prototype VXA archiver and virtual machine
are left for the next section.


\subsection{Trends and Design Principles}

Archived data is almost always compressed in some fashion to save space.
The one-time cost of compressing the data in the first place is usually
well justified by the savings in storage costs
(and perhaps network bandwidth)
offered by compression over the long term.

A basic property of data compression,
however,
is that the more you know about the data being compressed,
the more effectively you can compress it.
General string-oriented compressors such as {\tt gzip}
do not perform well on digitized photographs, audio, or video,
because the information redundancy present in digital media
does not predominantly take the form of repeated byte strings,
but is specific to the type of media.
For this reason a wide variety of media-specific compressors
have appeared recently.
{\em Lossless} compressors
achieve moderate compression ratios
while preserving all original information content,
while {\em lossy} compressors
achieve higher compression ratios
by discarding information whose loss is deemed ``unlikely to be missed''
based on semantic knowledge of the data.
Specialization of compression algorithms
is not limited to digital media:
compressors for semistructured data
such as XML are also available for example~\cite{liefke99xmill}.
This trend toward specialized encodings
leads to a first important design principle
for efficient archival storage:

\begin{quote}
{\em An archival storage system must permit use of
multiple, specialized compression algorithms.}
\end{quote}

Strong economic demand
for ever more sophisticated and effective data compression
has led to a rapid evolution in encoding schemes,
even within particular domains such as audio or video,
often yielding an abundance of mutually-incompatible competing schemes.
Even when open standards achieve widespread use,
the dominant standards evolve over time:
e.g., from Unix {\tt compress} to {\tt gzip} to {\tt bzip2}.
This trend leads to
VXA's second basic design principle:

\begin{quote}
{\em An archival storage system
must permit its set of compression algorithms
to evolve regularly.}
\end{quote}

The above two trends unfortunately work against
the basic purpose of archival storage:
to store data so that it remains available and usable later,
perhaps decades later.
Even if data is always archived using the latest encoding software,
that software---and the operating systems it runs on---%
may be long obsolete a few years later when the archived data is needed.
The widespread use of lossy encoding schemes compounds this problem,
because periodically decoding and re-encoding archived data
using the latest schemes would cause progressive information loss
and thus is not generally a viable option.
This constraint leads to VXA's third basic design principle:

\begin{quote}
{\em Archive extraction must be possible
without specific knowledge of the data's encoding.}
\end{quote}

VXA satisfies these constraints
by storing executable decoders with all archived data,
and by ensuring that these decoders run in a simple, well-defined, portable,
and thus hopefully relatively ``future-proof'' virtual environment.

\subsection{Creating Archives}

\begin{figure}[t]
\centerline{\epsfig{file=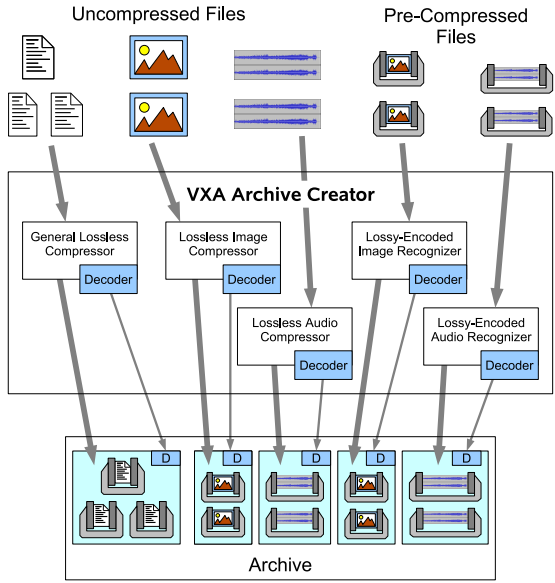, scale=0.30}}
\caption{Archive Writer Operation}
\label{fig-writer}
\end{figure}

Figure~\ref{fig-writer} illustrates the basic structure
of an archive writer in the VXA architecture.
The archiver contains a number of encoder/decoder or {\em codec} pairs:
several specialized codecs designed to handle specific content types
such as audio, video, or XML,
and at least one general-purpose lossless codec.
The archiver's codec set is extensible via plug-ins,
allowing the use of specialized codecs
for domain-specific content when desired.

The archiver accepts both uncompressed and already-compressed files as inputs,
and automatically tries to compress previously uncompressed input files
using a scheme appropriate for the file's type if available.
The archiver attempts to compress files of unrecognized type
using a general-purpose lossless codec such as {\tt gzip}.
By default the archiver uses only lossless encoding schemes
for its automatic compression,
but it may apply lossy encoding
at the specific request of the operator.

The archiver writes into the archive
a copy of the decoder portion of each codec it uses to compress data.
The archiver of course needs to include
only one copy of a given decoder in the archive,
amortizing the storage cost of the decoder
over all archived files of that type.

The archiver's codecs can also recognize
when an input file is {\em already} compressed in a supported format.
In this case,
the archiver just copies the pre-compressed data into the archive,
since re-compressing already-compressed data is generally ineffective
and particularly undesirable when lossy compression is involved.
The archiver still includes a copy of the appropriate decoder in the archive,
ensuring the data's continuing usability
even after the original codec has become obsolete or unavailable.

Some of the archiver's codecs
may be incapable of compression,
but may instead merely recognize files already encoded
using other, standalone compressors,
and attach a suitable decoder to the archived file.
We refer to such pseudo-codecs as {\em recognizer-decoders},
or {\em redecs}.

\subsection{Reading Archives}

\begin{figure}[t]
\centerline{\epsfig{file=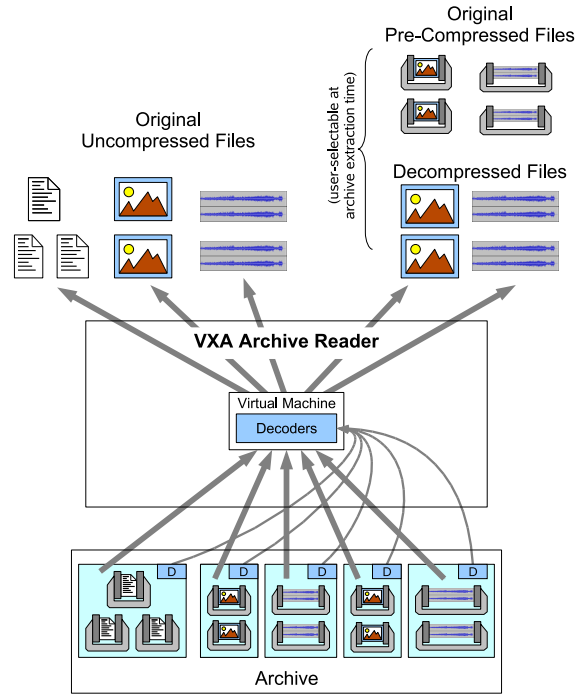, scale=0.30}}
\caption{Archive Reader Operation}
\label{fig-reader}
\end{figure}

Figure~\ref{fig-reader} illustrates the basic structure
of the VXA archive reader.
Unlike the writer,
the reader does not require a collection of content-specific codecs,
since all the decoders it needs are embedded in the archive itself.
Instead, the archive reader implements a virtual machine
in which to run those archived decoders.
To decode a compressed file in the archive,
the archive reader first locates the associated decoder in the archive
and loads it into its virtual machine.
The archive reader then executes the decoder in the virtual machine,
supplying the encoded data to the decoder
while accepting decoded data from the decoder,
to produce the decompressed output file.

The archive reader by default only decompresses files
that weren't already compressed when the archive was written.
This way, archived files that were already compressed
in popular standard formats such as JPEG or MP3,
which tend to be widely and conveniently usable in their compressed form,
remain compressed by default after extraction.
The reader can, however, be forced
to decode {\em all} archived files having an associated decoder,
as shown in Figure~\ref{fig-reader},
ensuring that encoded data remains decipherable
even if ``native'' decoders for the format disappear.

This capability also helps protect against data corruption
caused by codec bugs or evolution of standards.
If an archived audio file was generated by a buggy MP3 encoder,
for example,
it may not play properly later under a different MP3 decoder
after extraction from the archive in compressed form.
As long as the audio file was originally archived
with the specific (buggy) MP3 decoder that can decode the file correctly,
however,
the archive reader can still be instructed to use that archived decoder
to recover a usable decompressed audio stream.

The VXA archive reader
does not {\em always} have to use the archived x86-based decoders
whenever it extracts files from an archive.
To maximize performance,
the reader might by default
recognize popular compressed file types
and decode them using non-virtualized decoders
compiled for the native host architecture.
Such a reader
would fall back on running a virtualized decoder from the archive
when no suitable native decoder is available,
when the native decoder does not work properly
on a particular archived stream,
or when explicitly checking the archive's integrity.
Even if the archive reader commonly
uses native rather than virtualized decoders,
the presence of the VXA decoders in the archive
provides a crucial long-term fallback path for decoding,
ensuring that the archived information remains decipherable
after the codec it was compressed with
has become obsolete and difficult to find.

Routinely using native decoders to read archives
instead of the archived VXA decoders,
of course,
creates the important risk that a bug in a VXA decoder
might go unnoticed for a long time,
making an archive seem work fine in the short term
but be impossible to decode later
after the native decoder disappears.
For this reason,
it is crucial that explicit archive integrity tests
always run the archived VXA decoder,
and in general it is safest if the archive reader
always uses the VXA decoder even when native decoders are available.
Since users are unlikely to adopt this safer operational model consistently
unless VXA decoder efficiency is on par with native execution,
the efficiency of decoder virtualization
is more important in practice than it may appear in theory.

\subsection{The VXA Virtual Machine}

The archive reader's virtual machine isolates the decoders it runs
from both the host operating system and the processor architecture
on which the archive reader itself runs.
Decoders running in the VXA virtual machine 
have access to the computational primitives of the underlying processor
but are extremely limited in terms of input/output.
The only I/O decoders are allowed
is to read an encoded data stream supplied by the archive reader
and produce a corresponding decoded output stream.
Decoders cannot access any host operating system services,
such as to open files, communicate over the network,
or interact with the user.

Through this strong isolation,
the virtual machine not only ensures that decoders remain generic
and portable across many generations of operating systems,
but it also protects the host system
from buggy or malicious decoders that may be embedded in an archive.
Assuming the virtual machine is implemented correctly,
the worst harm a decoder can cause
is to garble the data it was supposed to produce
from a particular encoded file.
Since a decoder cannot communicate,
obtain information about the host system,
or even check the current system time,
decoders do not have access to information
with which they might deliberately ``sabotage'' their data
based on the conditions under which they are run.

When an archive contains many files associated with the same decoder,
the archive reader has the option of
re-initializing the virtual machine
with a pristine copy of the decoder's executable image
before processing each new file,
or reusing the virtual machine's state
to decode multiple files in succession.
Reusing virtual machine state may improve performance,
especially on archives containing many small files,
at the cost of introducing the risk
that a buggy or malicious decoder might ``leak'' information
from one file to another during archive extraction,
such as from a sensitive password or private key file
to a multimedia stream that is likely to appear on a web page.
The archive reader can minimize this security risk in practice
by always re-initializing the virtual machine
whenever the security attributes of the files it is processing change,
such as Unix owner/group identifiers and permissions.

The VXA virtual machine is based on the standard 32-bit x86 architecture:
all archived decoder executables are represented as x86-32 code,
regardless of the actual processor architecture of the host system.
The choice of the ubiquitous x86-32 architecture
ensures that almost any existing decoder written in any language
can be easily ported to run on the VXA virtual machine.

Although continuous improvements in processor hardware
are likely to make the performance of an archived VXA decoder
largely irrelevant over the long term,
compressed archives are frequently used for more short-term purposes as well,
such as making and restoring backups,
distributing and installing software,
and packaging XML-based structured documents~\cite{openoffice-xml}.
Archive extraction performance is crucial to these short-term uses,
and an archival storage system that performs poorly now
is unlikely to receive widespread adoption
regardless of its long-term benefits.
Besides supporting the re-use of existing decoder implementations,
VXA's adoption of the x86 architecture
also enables those decoders to run quite efficiently
on x86-based host processors,
as demonstrated later in Section~\ref{sec-eval}.
Implementing the VM efficiently on other architectures
requires binary translation,
which is more difficult and may be less efficient,
but is nevertheless by now a practical and proven
technology~\cite{sites93binary, chernoff98fx32, dehnert03transmeta, baraz03ia}.

\subsection{Applicability}

The VXA architecture
does not address the complete problem
of preserving the long-term usability of archived digital information.
The focus of VXA is on preserving {\em compressed} data streams,
for which simpler uncompressed formats are readily available
that can represent the same information.
VXA will not necessarily help with old proprietary word processor documents,
for example,
for which there is often no obvious ``simpler form''
that preserves all of the original semantic information.

Many document processing applications, however,
are moving toward use of
``self-describing'' XML-based structured
data formats~\cite{openoffice-xml},
combined with
a general-purpose ``compression wrapper'' such as ZIP~\cite{info-zip}
for storage efficiency.
The VXA architecture may benefit the compression wrapper in such formats,
allowing applications to encode documents
using proprietary or specialized algorithms for efficiency
while preserving the interoperability benefits of XML.
VXA's support for specialized compression schemes
may be particularly important for XML,
in fact,
since ``raw'' XML is extremely space-inefficient
but can be compressed most effectively
given some specialized knowledge of the data~\cite{liefke99xmill}.


\section{Archiver Implementation}
\label{sec-impl}

Although the basic VXA architecture as described above
could be applied to many archival storage or backup systems,
the prototype implementation explored in this paper
takes the form of an enhancement to
the venerable ZIP/UnZIP archival tools~\cite{info-zip}.
The ZIP format was chosen
over the {\tt tar}/{\tt gzip} format popular on Unix systems
because ZIP compresses files individually
rather than as one continuous stream,
making it amenable to treating files of different types
using different encoders.

For clarity,
we will refer to the new VXA-enhanced ZIP and UnZIP utilities here
as \vxzip{} and \vxunz{},
and to the modified archive format as ``vxZIP format.''
In practice, however,
the new tools and archive format can be treated as
merely a natural upgrade to the existing ones.

\subsection{ZIP Archive Format Modifications}

The enhanced vxZIP archive format retains
the same basic structure and features as the existing ZIP format,
and the new utilities remain backward compatible
with archives created with existing ZIP tools.
Older ZIP tools can list the contents
of archives created with \vxzip{},
but cannot extract files
requiring a VXA decoder.


The ZIP file format historically uses
a relatively fixed, though gradually growing, collection
of general-purpose lossless codecs,
each identified by a ``compression method'' tag
in a ZIP file.
A particular ZIP utility generally compresses all files
using only one algorithm by default---%
the most powerful algorithm it supports---%
and UnZIP utilities include built-in decoders
for most of the compression schemes used by past ZIP utilities.
(Decoders for the old LZW-based ``shrinking'' scheme
were commonly omitted for many years
due to the LZW patent~\cite{battilana04gif},
illustrating one of the practical challenges
to preserving archived data usability.)

\begin{figure}[t]
\centerline{\epsfig{file=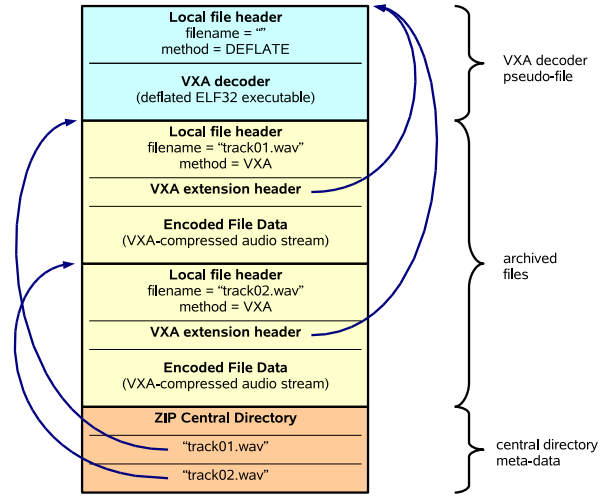, scale=0.30}}
\caption{vxZIP Archive Structure}
\label{fig-format}
\end{figure}

In the enhanced vxZIP format,
an archive may contain files compressed
using a mixture of traditional ZIP compression methods
and new VXA-specific methods.
Files archived using traditional methods
are assigned the standard method tag,
permitting even VXA-unaware UnZIP tools
to identify and extract them successfully.
The vxZIP format reserves one new ``special'' ZIP method tag
for files compressed using VXA codecs
that do not have their own ZIP method tags,
and which thus can only be extracted
with the help of an attached VXA decoder.

Regardless of whether an archived file
uses a traditional or VXA compression scheme,
\vxzip{} attaches a new VXA extension header to each file,
pointing to the file's associated VXA decoder,
as illustrated in Figure~\ref{fig-format}.
Using this extension header,
a VXA-aware archive reader can decode any archived file
even if it has an unknown method tag.
At the same time,
\vxunz{} can still use a file's ZIP method tag
to recognize files compressed using well-known algorithms
for which it may have a faster native decoder.

When \vxzip{} recognizes an input file
that is already compressed using a scheme
for which it has a suitable VXA decoder,
it stores the pre-compressed file directly without further compression
and tags the file with compression method 0 (no compression).
This method tag indicates to \vxunz{}
that the file should normally be left compressed on extraction,
and enables older UnZIP utilities
to extract the file in its original compressed form.
The \vxzip{} archiver nevertheless attaches a VXA decoder to the file
in the same way as for automatically-compressed files,
so that \vxunz{} can later be instructed
to decode the file all the way to its uncompressed form
if desired.

\subsection{Archiving VXA Decoders}

Since the 64KB size limitation of ZIP extension headers
precludes storing VXA decoders themselves in the file headers,
\vxzip{} instead stores each decoder
elsewhere in the archive as a separate ``pseudo-file''
having its own local file header
and an empty filename.
The VXA extension headers attached to ``actual'' archived files
contain only the ZIP archive offset of the decoder pseudo-file.
Many archive files can thus refer to one VXA decoder
merely by referring to the same ZIP archive offset.

ZIP archivers write a {\em central directory}
to the end of each archive,
which summarizes the filenames and other meta-data
of all files stored in the archive.
The \vxzip{} archiver includes entries in the central directory
only for ``actual'' archived files,
and not for the pseudo-files containing archived VXA decoders.
Since UnZIP tools normally use the central directory
when listing the archive's contents,
VXA decoder pseudo-files do not show up in such listings
even using older VXA-unaware UnZIP tools,
and old tools can still use the central directory
to find and extract any files
not requiring VXA-specific decoders.

A VXA decoder itself
is simply an ELF executable
for the 32-bit x86 architecture~\cite{tis95elf},
as detailed below in Section~\ref{sec-vm}.
VXA decoders are themselves compressed in the archive
using a fixed, well-known algorithm:
namely the ubiquitous ``deflate'' method
used by existing ZIP tools
and by the {\tt gzip} utility popular on Unix systems.

\subsection{Codecs for the Archiver}

Since a basic goal of the VXA architecture
is to be able to support a wide variety of often specialized codecs,
it is unacceptable for \vxzip{}
to have a fixed set of built-in compressors,
as was generally the case for previous ZIP tools.
Instead,
\vxzip{} introduces a plug-in architecture
for codecs to be used with the archiver.
Each codec consists of two main components:

\begin{itemize}
\item	The encoder is a standard dynamic-link library (DLL),
	which the archiver loads into its own address space at run-time,
	and invokes directly to recognize and compress files.
	The encoder thus runs ``natively'' on the host processor architecture
	and in the same operating system environment as the archiver itself.
\item	The decoder is an executable image for the VXA virtual machine,
	which the archiver writes into the archive
	if it produces or recognizes any encoded files using this codec.
	The decoder is always an ELF executable for the 32-bit x86 architecture
	implemented by the VXA virtual machine,
	regardless of the host processor architecture and operating system
	on which the archiver actually runs.
\end{itemize}

A natural future extension to this system
would be to run VXA encoders as well as decoders
in a virtual machine,
making complete codec pairs maximally portable.
While such an extension should not be difficult,
several tradeoffs are involved.
A virtual machine for VXA encoders
may require user interface support
to allow users to configure encoding parameters,
introducing additional system complexity.
While the performance impact of the VXA virtual machine is not severe
at least on x86 hosts,
as demonstrated in Section~\ref{sec-eval},
implementing encoders as native DLLs
enables the archiving process to run with maximum performance
on any host.
Finally,
vendors of proprietary codecs
may not wish to release their encoders
for use in a virtualized environment,
because it might make license checking more difficult.
For these reasons,
virtualized VXA encoders are left for future work.

\section{The Virtual Machine}
\label{sec-vm}

The most vital component of the \vxunz{} archive reader
is the virtual machine in which it runs archived decoders.
This virtual machine is implemented
by \vx32{}, a novel {\em virtual machine monitor} (VMM)
that runs in user mode
as part of the archive reader's process,
without requiring any special privileges
or extensions to the host operating system.
Decoders under \vx32{} effectively run within \vxunz's address space,
but in a software-enforced fault isolation
domain~\cite{wahbe93efficient},
protecting the application process
from possible actions of buggy or malicious decoders.
The VMM is implemented
as a shared library linked into \vxunz{};
it can also be used to implement specialized virtual machines
for other applications.

The \vx32{} VMM
currently runs only on x86-based host processors,
in both 32-bit and the new 64-bit modes.
The VMM relies on quick x86-to-x86 
code scanning and translation techniques
to sandbox a decoder's code as it executes.
These techniques are comparable to those
used by Embra~\cite{witchel96embra},
VMware~\cite{sugarman01virtualizing},
and Valgrind~\cite{nethercote03valgrind},
though \vx32{} is simpler as it need only provide isolation,
and not simulate a whole physical PC
or instrument object code for debugging.
Full binary translation
to make \vx32{} run on other host architectures is under development.

\subsection{Data Sandboxing}

The VXA virtual machine provides decoders
with a ``flat'' unsegmented address space up to 1GB in size,
which always starts at virtual address 0
from the perspective of the decoder.
The VM does not allow decoders access to the underlying x86 architecture's
legacy segmentation facilities.
The \vx32{} VMM does, however,
{\em use} the legacy segmentation features of the x86 host processor
in order to implement the virtual machine efficiently.

\begin{figure}[t]
\centerline{\epsfig{file=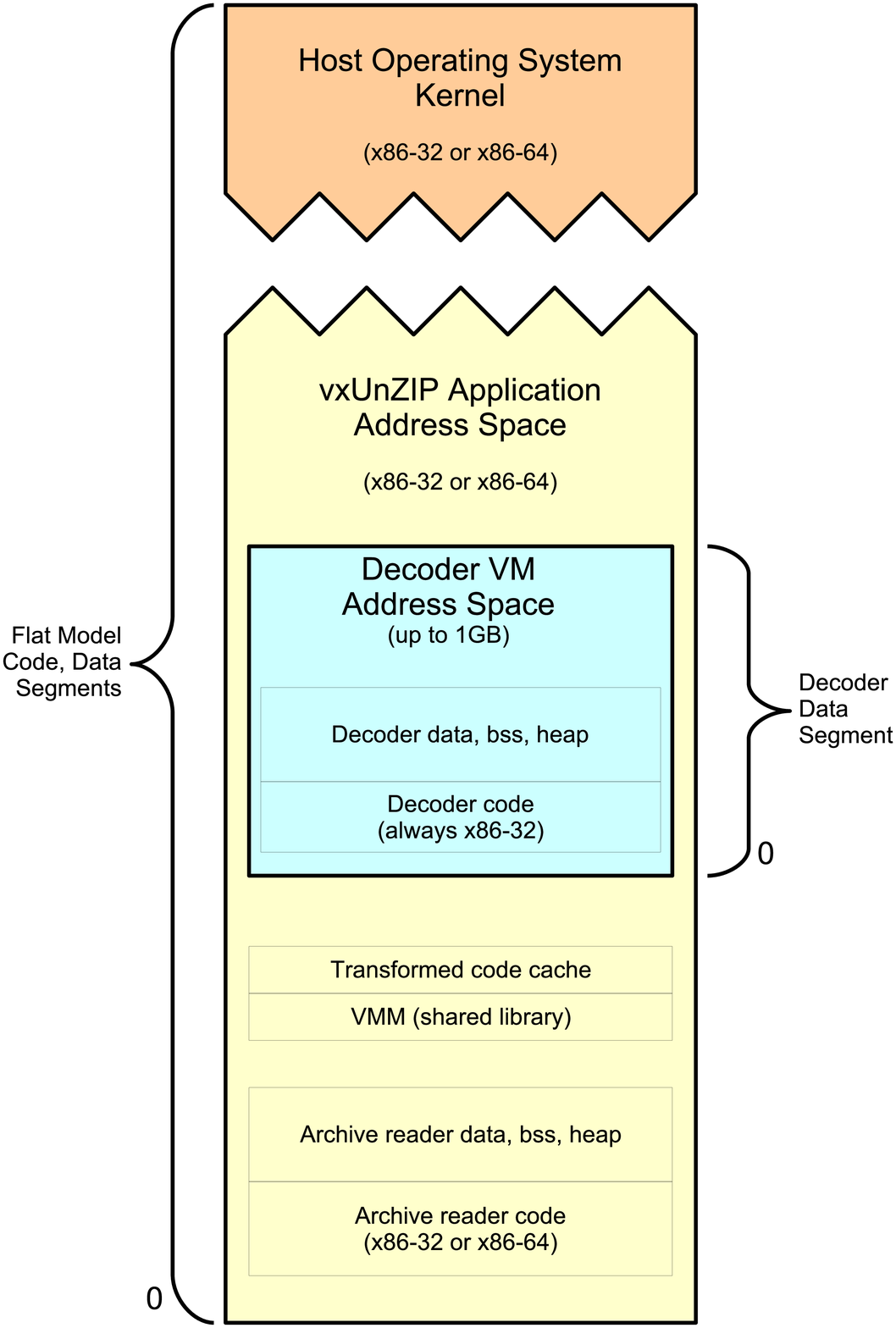, scale=0.30}}
\caption{Archive Reader and VMM Address Spaces}
\label{fig-vm}
\end{figure}

As illustrated in Figure~\ref{fig-vm},
\vx32{} maps a decoder's virtual address space
at some arbitrary location within its own process,
and sets up a special process-local (LDT) data segment
with a base and limit that provides access only to that region.
While running decoder code,
the VMM keeps this data segment
loaded into the host processor's segment registers
that are used for normal data reads and writes (DS, ES, and SS).
The decoder's computation and memory access instructions
are thus automatically restricted to the sandbox region,
without requiring the special code transformations
needed on other architectures~\cite{wahbe93efficient}.
\com{
Since computation and data read/write instructions
usually form the bulk of a codec's instruction stream,
this use of the host processor's segmentation facility
provides an important performance benefit.
}

Although the legacy segmentation features that the VMM depends on
are not functional
in the 64-bit addressing mode (``long mode'') of the new x86-64 processors,
these processors
provide 64-bit applications the ability
to switch back to a 32-bit ``compatibility mode''
in which segmentation features are still available.
On a 64-bit system,
\vxunz{} and the VMM run in 64-bit long mode,
but decoders run in 32-bit compatibility mode.
Thus, \vx32{} runs equally well
on both x86-32 and x86-64 hosts
with only minor implementation differences in the VMM
(amounting to about 100 lines of code).

\subsection{Code Sandboxing}
\label{sec-code-sandbox}

Although the VMM could similarly set up an x86 code segment
that maps only the decoder's address space,
doing so would not by itself prevent decoders
from executing arbitrary x86 instructions
that are ``unsafe'' from the perspective of the VMM,
such as those that would modify the segment registers
or invoke host operating system calls directly.
On RISC-based machines with fixed instruction sizes,
a software fault isolation VMM
can solve this problem by scanning the untrusted code
for ``unsafe'' code sequences
when the code is first loaded~\cite{wahbe93efficient}.
This solution is not an option
on the x86's variable-length instruction architecture,
unfortunately,
because within a byte sequence comprising one or more legitimate instructions
there may be sub-sequences forming unsafe instructions,
to which the decoder code might jump directly.
The RISC-based techniques
also reserve up to five general-purpose registers
as {\em dedicated registers} to be used for fault isolation,
which is not practical on x86-32
since the architecture provides only eight general-purpose registers total.

The \vx32{} VMM therefore never executes decoder code directly,
but instead dynamically scans decoder code sequences to be executed
and transforms them into ``safe'' code fragments
stored elsewhere in the VMM's process.
As with Valgrind~\cite{nethercote03valgrind}
and just-in-time compilation
techniques~\cite{deutsch84efficient, krall98efficient},
the VMM keeps transformed code fragments in a cache
to be reused whenever the decoder
subsequently jumps to the same virtual entrypoint again.

The VMM must of course transform all flow control instructions
in the decoder's original code
so as to keep execution confined to the safe, transformed code.
The VMM rewrites branches with fixed targets
to point to the correct transformed code fragment if one already exists.
Branches to fixed but as-yet-unknown targets
become branches to a ``trampoline'' that, when executed,
transforms the target code
and then back-patches the original (transformed) branch instruction
to point directly to the new target fragment.
Finally, the VMM rewrites indirect branches
whose target addresses are known only at runtime
(including function return instructions),
so as to look up the target address dynamically
in a hash table of transformed code entrypoints.

\subsection{Virtual System Calls}

The \vx32{} VMM rewrites x86 instructions
that would normally invoke system calls to the host operating system,
so as to return control to the user-mode VMM instead.
In this way,
\vx32{} ensures that decoders have no direct access to host OS services,
but can only make controlled ``virtual system calls''
to the VMM or the archive reader.

Only five virtual system calls are available
to decoders running under \vxunz{}:
{\tt read}, {\tt write}, {\tt exit}, {\tt setperm}, and {\tt done}.
The first three have their standard Unix meanings,
while {\tt setperm} supports heap memory allocation,
and {\tt done} enables decoders to signal to \vxunz{}
that they have finished decoding one stream
and are able to process another without being re-loaded.
Decoders have access to
three standard ``virtual file handles''---%
{\tt stdin}, {\tt stdout}, and {\tt stderr}---%
but have no way to open any other files.
A decoder's virtual {\tt stdin} file handle
represents the data stream to be decoded,
its {\tt stdout}
is the data stream it produces by decoding the input,
and {\tt stderr}
serves the traditional purpose of allowing the decoder
to write error or debugging messages.
(\vxunz{} only displays such messages from decoders
when in verbose mode.)
A VXA decoder is therefore a traditional Unix filter
in a very pure form.

Since a decoder's address space
comprises a portion of \vxunz's own address space,
the archive reader can easily access the decoder's data directly
for the purpose of servicing virtual system calls,
in the same way that the host OS kernel services system calls
made by application processes.
To handle the decoder's {\tt read} and {\tt write} calls,
\vxunz{} merely passes the system call on to the native host OS
after checking and adjusting the file handle and buffer pointer arguments.
A decoder's I/O calls thus require no extra data copying,
and the indirection through the VMM and \vxunz{} code is cheap
as it does not cross any hardware protection domains.

\section{Evaluation and Results}
\label{sec-eval}

This section experimentally evaluates
the prototype \vxzip/\vxunz{} tools
in order to analyze the practicality of the VXA architecture.
The two most obvious questions
about the practicality of VXA
are whether running decoders in a virtual machine
seriously compromises their performance
for short-term uses of archives
such as backups and software/data packaging,
and whether embedding decoders in archives
entails a significant storage cost.
We also consider the portability issues
of implementing virtual machines that run x86-32 code on other hosts.

\subsection{Test Decoders}

The prototype \vxzip{} archiver
includes codecs for several well-known compressed file formats,
summarized in Table~\ref{tab-decoders}.
The two general-purpose codecs, {\tt zlib} and {\tt bzip2},
are for arbitrary data streams:
\vxzip{} can use either of them as its ``default compressor''
to compress files of unrecognized type while archiving.
The remaining codecs are media-specific.
All of the codecs are based directly
on publicly-available libraries written in C,
and were compiled using a basic GCC
cross-compiler setup.

\begin{table*}
\begin{small}
\begin{center}
\begin{tabular}{l||l|l|l|}
Decoder		& Description	& Availability		& Output Format \\
\hline
\hline
\multicolumn{4}{c}{\bf General-Purpose Codecs} \\
{\tt zlib}	& ``Deflate'' algorithm from ZIP/gzip
				& {\tt www.zlib.net}	& (raw data)	\\
{\tt bzip2}	& Popular BWT-based algorithm
				& {\tt www.bzip.org}	& (raw data)	\\
\hline
\multicolumn{4}{c}{\bf Still Image Codecs} \\
{\tt jpeg}	& Independent JPEG Group (IJG) reference decoder
				& {\tt www.ijg.org}	& BMP image \\
{\tt jp2}	& JPEG-2000 reference decoder from JasPer library
				& {\tt www.jpeg.org/jpeg2000}
							& BMP image \\
\hline
\multicolumn{4}{c}{\bf Audio Codecs} \\
{\tt flac}	& Free Lossless Audio Codec (FLAC) decoder
				& {\tt flac.sourceforge.net}
							& WAV audio \\
{\tt vorbis}	& Ogg Vorbis audio decoder
				& {\tt www.vorbis.com}	& WAV audio \\
\hline
\end{tabular}
\end{center}
\end{small}
\caption{Decoders Implemented in \vxzip/\vxunz{} Prototype}
\label{tab-decoders}
\end{table*}

The {\tt jpeg} and {\tt jp2} codecs 
are recognizer-decoders (``redecs''),
which recognize still images compressed
in the lossy JPEG~\cite{wallace91jpeg}
and JPEG-2000~\cite{iso00jpeg} formats, respectively,
and attach suitable VXA decoders to archived images.
These decoders,
when run under \vxunz{},
output uncompressed images
in the simple and universally-understood Windows BMP file format.
The {\tt vorbis} redec
similarly recognizes compressed audio streams
in the lossy Ogg/Vorbis format~\cite{xiph04vorbis},
and attaches a Vorbis decoder
that yields an uncompressed audio file
in the ubiquitous Windows WAV audio file format.

Finally, {\tt flac} is a full encoder/decoder pair
for the Free Lossless Audio Codec (FLAC) format~\cite{coalson05flac}.
Using this codec,
\vxzip{} can not only recognize audio streams
already compressed in FLAC format and attach a VXA decoder,
but it can also recognize {\em uncompressed} audio streams in WAV format
and automatically compress them using the FLAC encoder.
This codec thus demonstrates how a VXA archiver
can make use of compression schemes specialized to particular types of data,
without requiring the archive reader to contain built-in decoders
for each such specialized compression scheme.

The above codecs with widely-available open source implementations
were chosen for purposes of {\em evaluating}
the prototype \vxzip/\vxunz{} implementation,
and are not intended to serve as ideal examples
to {\em motivate} the VXA architecture.
While the open formats above may gradually evolve over time,
their open-source decoder implementations are unlikely to disappear soon.
Commercial archival and multimedia compression products
usually incorporate proprietary codecs,
however,
which might serve as better ``motivating examples'' for VXA:
proprietary codecs tend to evolve more quickly
due to intense market pressures,
and and their closed-source implementations
cannot be maintained by the customer
or ported to new operating systems
once the original product is obsolete and unsupported by the vendor.

\subsection{Performance of Virtualized Decoders}

\begin{figure*}[t]
\centerline{\epsfig{file=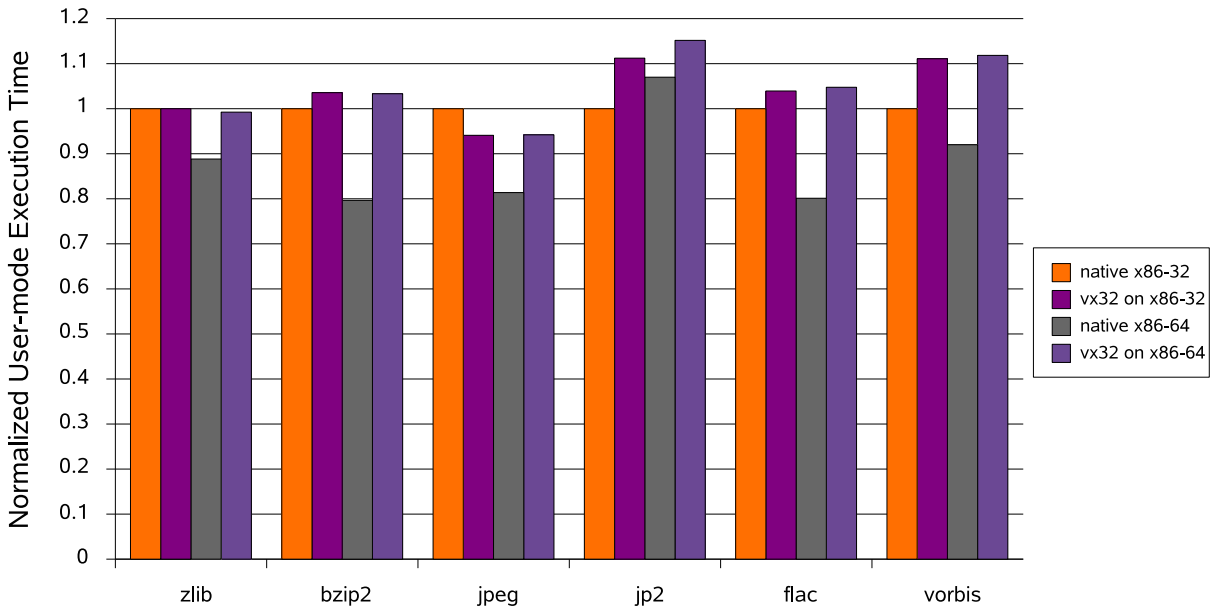, scale=0.8}}
\caption{Performance of Virtualized Decoders}
\label{fig-perf}
\end{figure*}

\com{	Some raw data:

	x86-32		x86-64
	host	vx32	host	vx32
zlib	 1.569	 1.654	 1.272	 1.655
bz2	13.251	14.062	11.207	14.161
flac	 2.723	 2.866	 2.230	 2.958
}

To evaluate the performance cost of virtualization,
the graph in Figure~\ref{fig-perf}
shows the user-mode CPU time consumed
running the above decoders over several test data sets,
both natively and under the \vx32{} VMM.
All execution times are normalized
to the native execution time on an x86-32 host system.
The data set used to test the general-purpose lossless codes
is a Linux 2.6.11 kernel source tree;
the data sets used for the media-specific codecs
consist of typical pictures and music files
in the appropriate format.
All tests were run on an AMD Athlon 64 3000+
with 512MB of RAM,
on both the x86-32 and x86-64 versions of SuSE Linux 9.3.
The same compiler version (GCC 4.0.0) and optimization settings ({\tt -O3})
were used for the native and virtualized versions of each decoder,
and the timings represent user-mode process time
as reported by the {\tt time} command
so as to factor out disk and system overhead.
Total wall-clock measurements are not shown
because for all but the slowest decoder, jp2,
disk overhead dominates total wall-clock time
and introduces enough additional variance between successive runs
to swamp the differences in CPU-bound decoding time.

\widowpage
As Figure~\ref{fig-perf} shows,
the decoders running under the \vx32{} VMM
experience a slowdown of up to 11\%
relative to native x86-32 execution.
The {\tt vorbis} decoder initially experienced a 29\% slowdown
when compiled for VXA unmodified,
due to subroutine calls
in the decoder's inner loop
that accentuate the VMM's flow-control overhead
by requiring hash table lookups (see Section~\ref{sec-code-sandbox}).
Inlining these two functions
both improved the performance of the native decoder slightly (about 1\%)
and reduced the relative cost of virtualization to 11\%.
The other decoders were unmodified from their original distribution form.
The JPEG decoder became slightly faster under \vx32,
possibly due to effects of the VMM's code rewriting
on instruction cache locality;
such effects are possible
and have been exploited elsewhere~\cite{bala00dynamo}.

The virtualized decoders fall farther behind
in comparison with native execution on an x86-64 host,
but this difference is mostly due
to the greater efficiency of the 64-bit native code
rather than to virtualization overhead.
Virtualized decoders always run in 32-bit mode
regardless of the host system,
so their absolute performance is almost identical
on 32-bit versus 64-bit hosts,
as the graph shows.

\subsection{Decoder Storage Overhead}

\begin{table*}
\begin{center}
\begin{tabular}{l||r|rr|rr||r}
Decoder		& \multicolumn{5}{|c||}{Code Size}
				& \multicolumn{1}{c}{Compressed} \\
		& Total	& \multicolumn{2}{|c|}{Decoder}
				& \multicolumn{2}{|c||}{C Library}
				& \multicolumn{1}{c}{({\tt zlib})} \\
\hline
\hline
\multicolumn{6}{c}{\bf General-Purpose Codecs} \\
{\tt zlib}	&46.0KB	& 32.4KB & (70\%)	& 13.6KB & (30\%) & 26.2KB \\
{\tt bzip2}	&71.1KB	& 60.9KB & (86\%)	& 10.2KB & (14\%) & 29.9KB \\
\hline
\multicolumn{6}{c}{\bf Still Image Codecs} \\
{\tt jpeg}	&103.3KB& 90.0KB & (87\%)	& 13.3KB & (13\%) & 48.6KB \\
{\tt jp2}	&220.2KB& 198.5KB & (90\%)	& 21.7KB & (10\%) & 105.9KB \\
\hline
\multicolumn{6}{c}{\bf Audio Codecs} \\
{\tt flac}	&102.5KB& 84.2KB & (82\%)	& 18.3KB & (18\%) & 47.6KB \\
{\tt vorbis}	&233.4KB& 200.3KB & (86\%)	& 33.1KB & (14\%) & 129.7KB \\
\hline
\end{tabular}
\end{center}
\caption{Code Size of Virtualized Decoders}
\label{tab-size}
\end{table*}

To evaluate the storage overhead of embedding decoders in archives,
Table~\ref{tab-size} summarizes 
the size of each decoder's executable image
when compiled and linked for the VXA virtual machine.
The code size for each decoder is further split
into the portion comprising the decoder itself
versus the portion derived from the statically-linked C library
against which each decoder is linked.
No special effort was made to trim unnecessary code,
and the decoders were compiled to optimize performance
over code size.

The significance of these absolute storage overheads
of course
depends on the size of the archive in which they are embedded,
since only one copy of a decoder
needs to be stored in the archive
regardless of the number of encoded files that use it.
As a comparison point, however,
a single 2.5-minute CD-quality song
in the dataset used for the earlier performance tests,
compressed at 120Kbps using the lossy Ogg codec,
occupies 2.2MB.
The 130KB Ogg decoder for VXA therefore
represents a 6\% space overhead
in an archive containing only this one song,
or a 0.6\% overhead in an archive containing a 10-song album.
The same 2.5-minute song compressed using the lossless FLAC codec
occupies 24MB,
next to which the 48KB \vx32{} decoder
represents a negligible 0.2\% overhead.

\com{
\subsection{Correctness}

Testing reliability, determinism
across processor implementations:
random vx32 instruction streams
}

\subsection{Portability Considerations}


A clear disadvantage of using the native x86 processor architecture
as the basis for VXA decoders
is that porting the archive reader to non-x86 host architectures
requires instruction set emulation or binary translation.
While instruction set emulators can be quite portable,
they also tend to be many times slower than native execution,
making them unappealing for computation-intensive tasks
such as data compression.
Binary translation provides better performance
and has entered widespread commercial use,
but is not simple to implement,
and even the best binary translators are unlikely
to match the performance of natively-compiled code.

The QEMU x86 emulator~\cite{bellard05qemu}
introduces a binary translation technique
that offers a promising compromise between portability and performance.
QEMU uses a native C compiler for the host processor architecture
to generate short code fragments that emulate individual x86 instructions.
QEMU's dynamic translator then scans the x86 code at run-time
and pastes together the appropriate native code fragments
to form translated code.
While this method is unlikely to perform as well
as a binary translator designed and optimized
for a specific host architecture,
it provides a portable method
of implementing emulators that offer usable performance levels.

Even without efficient binary translation for x86 code,
however,
the cost of emulation
does not necessarily make the VXA architecture impractical
for non-x86 host architectures.
An archive reader
can still provide fast native decoders for currently popular file formats,
running archived decoders under emulation
only when no native decoder is available.
The resulting archival system
is no slower in practice than existing tools
based on a fixed set of compressors,
but provides the added assurance that archived data
will still be decipherable far into the future.
It is much better to be able to decode archived data slowly using emulation
than not to be able to decode it at all.

\subsection{Availability}

The \vxzip/\vxunz{} tools,
the \vx32{} virtual machine,
and the data sets used in the above tests
can be obtained from
\verb|http://pdos.csail.mit.edu/~baford/vxa/|.

\section{Related Work}
\label{sec-related}

The importance and difficulty of preserving digital information
over the long term
is gaining increasing recognition~\cite{garrett96preserving}.
This problem can be broken into two components:
preserving {\em data} and
preserving the data's {\em meaning}~\cite{crespo98archival}.
Important work is ongoing to address the first
aspect~\cite{goldberg98towards, cooper02peertopeer, maniatis05lockss},
and the second, the focus of this paper,
is beginning to receive serious attention.

\subsection{Archival Storage Strategies}

Storing executable decoders with archived data
is not new:
popular archivers including ZIP
often ship with tools to create {\em self-extracting archives},
or executables that decompress themselves when run~\cite{pkzip, info-zip}.
Such self-extracting archives are designed for convenience,
however,
and are traditionally specific to a particular host operating system,
making them as bad as or worse than traditional non-executable archives
for data portability and longevity.
Self-extracting archives also provide no security
against bugs or malicious decoders;
E-mail viruses routinely disguise themselves
as self-extracting archives supposedly containing useful applications.

Rothenberg suggested a decade ago the idea of
archiving the original application and system software used to create data
along with the data itself,
and using emulators to run archived software
after its original hardware platform
becomes obsolete~\cite{rothenberg95ensuring}.
Archiving entire systems
and emulating their hardware accurately is difficult, however,
because real hardware platforms (including necessary I/O devices)
are extremely complex
and tend to be only partly standardized and documented~\cite{bearman99reality}.
Preserving the {\em functionality} of the original system
is also not necessarily equivalent
to preserving the {\em usefulness} of the original data.
The ability to view old data in an emulator window
via the original application's archaic user interface,
for example,
is not the same as being able to load or ``cut-and-paste'' the data
into new applications or process it
using new indexing or analysis tools.

Lorie later proposed to archive data along with specialized decoder programs,
which run on a specialized ``Universal Virtual Computer'' (UVC),
and extract archived data into a self-describing
XML-like format~\cite{lorie00archiving}.
The UVC's simplicity makes emulation easier,
but since it represents a new architecture
substantially different from those of real processors,
UVC decoders must effectively be written from scratch in assembly language
until high-level languages and tools are developed~\cite{lorie02uvc}.
More importantly, the UVC's specialization
to the ``niche'' of long-term archival storage systems
virtually guarantees that
high-level languages, development tools, and libraries for it
will never be widely available or well-supported
as they are for general-purpose architectures.
\com{
Finally, the UVC design's
lack of concern for execution efficiency,
while justifiable in the context of long-term storage
on the assumption that computers will keep getting faster,
neglects the many important and often more economically compelling
short-term uses of digital archives.
}

\com{
VXA in contrast addresses
both the short-term and long-term uses of digital archives,
by following a design path
between Rothenberg's ``whole system'' strategy
and Lorie's UVC approach.
While entire hardware platforms
are complex, rapidly evolving, incompletely specified,
and hence difficult to emulate precisely,
the {\em processor architecture} component of real platforms---%
particularly the x86 instruction set at the core of standard PCs---%
is well-understood,
thoroughly documented by multiple processor vendors,
and has religiously maintained backwards code compatibility
over decades of processor evolution.
By retaining architectural compatibility with real hardware
while isolating archival decoders
from the more complex and rapidly-evolving system components
such as I/O devices and operating system interfaces,
VXA enables re-use of existing x86 development tools and code
and efficient decoder execution on existing hardware,
in addition to preserving the archived data over the long term.
}

The LOCKSS archival system
supports data format converter plug-ins
that transparently migrate data in obsolete formats to new formats
when a user accesses the data~\cite{rosenthal05transparent}.
Over time,
however,
actively maintaining converter plug-ins
for an ever-growing array of obsolete compressed formats may become difficult.
Archiving VXA decoders with compressed data {\em now}
ensures that future LOCKSS-style ``migrate-on-access'' converters
will only need to read common historical {\em uncompressed} formats,
such as BMP images or WAV audio files,
and not the far more numerous and rapidly-evolving compressed formats.
VXA therefore complements a ``migrate-on-access'' facility
by reducing the number and variety of source formats 
the access-time converters must support.

\com{
XXX 
Check out:
http://citeseer.ist.psu.edu/chervenak98protecting.html
http://tom.library.upenn.edu/pubs/thesis/
http://www.imaging.org/store/epub.cfm?abstrid=30296
http://www.imaging.org/store/epub.cfm?abstrid=30329
}

\subsection{Specialized Virtual Environments}

Virtual machines and languages
have been designed for many specialized purposes,
such as
printing~\cite{adobe99postscript},
boot loading~\cite{ieee94boot},
Web programming~\cite{gosling96java, lucco95omniware},
packet filters~\cite{mogul87packet}
and other OS extensions~\cite{small96comparison},
active networks~\cite{tennenhouse97survey},
active disks~\cite{riedel98active},
and grid computing~\cite{chang02trustless}.
In this tradition,
VXA could be appropriately described as
an architecture for ``active archives.''

Similarly, dynamic code scanning and translation
is widely used for purposes
such as migrating legacy applications across processor
architectures~\cite{sites93binary, chernoff98fx32, baraz03ia},
simulating complete hardware platforms~\cite{witchel96embra},
run-time code optimization~\cite{bala00dynamo},
implementing new processors~\cite{dehnert03transmeta},
and debugging compiled code~\cite{nethercote03valgrind, seward05valgrind}.
In contrast with the common ``retroactive'' uses
of virtual machines and dynamic translation
to ``rescue old code'' that no longer runs on the latest systems,
however,
VXA applies these technologies {\em proactively}
to preserve the long-term usability and portability of archived data,
{\em before} the code that knows how to decompress it becomes obsolete.

Most virtual machines designed to support safe application extensions
rely on type-safe languages such as Java~\cite{case96implementing}.
In this case,
the constraints imposed by the language
make the virtual machine
more easily portable across processor architectures,
at the cost of requiring all untrusted code to be written in such a language.
While just-in-time compilation~\cite{deutsch84efficient, krall98efficient}
has matured to a point where
type-safe languages perform adequately for most purposes,
some software domains
in which performance is traditionally perceived as paramount---%
such as data compression---%
remain resolutely attached to unsafe languages such as C and assembly language.
Advanced digital media codecs
also frequently take advantage of the SIMD extensions
of modern processors~\cite{intel05ia32},
which tend to be unavailable in type-safe languages.
The desire to support
the many widespread open and proprietary data encoding algorithms
whose implementations are only available in unsafe languages,
therefore,
makes type-safe language technology infeasible for the VXA architecture.

\subsection{Isolation Technologies}

The prototype \vx32{} VMM demonstrates
a simple and practical software fault isolation (SFI) strategy on the x86,
which achieves performance comparable to previous techniques
designed for on RISC architectures~\cite{wahbe93efficient},
despite the fact that the RISC-based techniques
are not easily applicable to the x86
as discussed in Section~\ref{sec-code-sandbox}.
RISC-based SFI,
observed to incur a 15--20\% overhead for full virtualization,
can be trimmed to 4\% overhead by sandboxing memory writes but not reads,
thereby protecting the host application
from active interference by untrusted code
but not from snooping.
Unfortunately, this weaker security model
is probably not adequate for VXA:
a functional but malicious decoder
for multimedia files likely to be posted on the Web,
for example,
could scan the archive reader's address space
for data left over from restoring sensitive files
such as passwords and private keys from a backup archive,
and surreptitiously leak that information
into the (public) multimedia output stream it produces.

The Janus security system~\cite{goldberg96secure}
runs untrusted ``helper'' applications in separate processes,
using hardware-based protection in conjunction with
Solaris's sophisticated process tracing facilities
to control the supervised applications' access to host OS system calls.
This approach is more portable across processor architectures than \vx32's,
but less portable across operating systems
since it relies on features currently unique to Solaris.
The Janus approach also does not enhance
the portability of the helper applications,
since it does not insulate them
from those host OS services they {\em are} allowed to access.

The L4 microkernel used an x86-specific segmentation trick
analogous to \vx32's data sandboxing technique
to implement fast IPC between small address spaces~\cite{liedtke95improved}.
A Linux kernel extension
similarly used segmentation and paging in combination
to give user-level applications a sandbox
for untrusted extensions~\cite{tzicker99integrating}.
This latter technique can provide each application
with only one virtual sandbox at a time,
however,
and it imposes constraints on the kernel's own use of x86 segments
that would make it impossible to grant use of this facility
to 64-bit applications on new x86-64 hosts.

\section{Conclusion}
\label{sec-conc}

The VXA architecture for archival data storage
offers a new and practical solution to the problem
of preserving the usability of digital content.
By including executable decoders in archives
that run on a simple and OS-independent virtual machine
based on the historically enduring x86 architecture,
the VXA architecture ensures that archived data
can always be decoded
into simpler and less rapidly-evolving uncompressed formats,
long after the original codec has become obsolete
and difficult to find.
The prototype \vxzip/\vxunz{} archiver for x86-based hosts
is portable across operating systems,
and decoders retain good performance when virtualized.

\subsection*{Acknowledgments}

The author wishes to thank Frans Kaashoek, Russ Cox, Maxwell Krohn,
and the anonymous reviewers for many helpful comments and suggestions.

\begin{footnotesize}
\bibliography{vxa}
\bibliographystyle{plain}
\end{footnotesize}


\end{document}